\documentstyle[12pt,epic]{article}
\begin{document}

\title{Models for Pedestrian Behavior}

\author{Dirk Helbing \\II. Institut f\"{u}r Theoretische Physik\\
Universit\"{a}t Stuttgart}
\maketitle

\section{Abstract}

The behavior of pedestrians shows certain regularities, which can be
described by quantitative (partly stochastic) models. The models are based
on the behavior of {\sl individual} pedestrians, which depends on the
pedestrian {\sl intentions} on one hand, and on the 
aspects of {\sl movement} on the other hand. 
\par
The pedestrian intentions concerning a sequence of destinations
are influenced first by the {\sl demand} for certain kinds of commodities,
second by the {\sl location of stores} selling these kinds of commodities,
third by the {\sl expenditures} (prices, ways, etc.)
to get the required commodities.
\par
The {\sl actual} pedestrian {\sl movement} starts and ends at special
{\sl city entry points} like bus stops, parking lots or metro stations.
It is guided by the pedestrian intentions, but is subject to {\sl deceleration
processes} and {\sl avoidance maneuvers} due to obstacles or pedestrians
who are in the way. As a consequence, the pedestrians have to speed up 
in the course of time in order
to reach the next destination well-timed. In addition,
the pedestrian behavior is influenced by unexpected {\sl attractions}
(e.g. by shop windows or entertainment). 
\par
The model for the behavior of individual pedestrians is an ideal starting
point for {\sl computer simulations} of pedestrian 
crowds. Such simulations
take into account the limited {\sl capacity} of pedestrian ways and places,
and allow to determine an optimal design of pedestrian
areas and an optimal arrangement of store locations. 
Therefore, they can be applied for {\sl town- and traffic-planning}.
\par
The model for the behavior of individual pedestrians also allows the derivation
of mathematical equations for {\sl pedestrian crowds} and for 
{\sl pedestrian groups}. Pedestrian crowds 
can be described by a {\sl stochastic formulation}, by a 
{\sl gaskinetic formulation} or by a {\sl fluiddynamic formulation}.
The gaskinetic formulation ({\sl mezoscopic level})
can be derived from the stochastic formulation ({\sl microscopic level}),
and the fluiddynamic formulation ({\sl macroscopic level})
from the gaskinetic formulation (mezoscopic level).

\section{Introduction}

Human behavior is based on individual decisions. In building a mathematical
model for the movement of pedestrians, one has to assume that these
decisions show certain regularities (e.g. follow {\sl stochastic laws}). 
This assumption
is justified, because decisions and the behavior of pedestrians are
usually determined by {\sl utility maximization}: 
For example, a pedestrian takes an optimal path to a chosen destination, and 
tries to minimize delays when having to avoid obstacles or other 
pedestrians. The
optimal behavior for a given situation can be derived by plausibility
considerations, and will be used as a model for pedestrian movement.
Of course this optimal behavior is normally not thought about,
but by {\sl trial and error} an individual has automatically learned to use
the most successful behavioral strategy, when being confronted with a 
standard situation.

\section{Individual behavior} \label{model}

The behavior of individual pedestrians is the (microscopic) basis for 
developing models that describe pedestrian groups or pedestrian crowds.
A model for the individual behavior has to take into account the pedestrian
intentions and the aspects of movement. In the
following, the basic ideas of a model of this kind will be described.
The {\sl mathematical} formulation of the
model will be presented in a forthcoming paper. 

\subsection{Pedestrian intentions}

Let us consider the case of pedestrians who walk in a {\sl shopping area}. 
(This is the most relevant case for town- and traffic-planning.) 
The pedestrians' behavior will
be determined mainly by their {\sl demand}, then. This demand will be varying
according to a certain distribution, which may depend on the pedestrians'
{\sl consumer type}. Given a certain demand, 
a pedestrian's {\sl destinations} will
be stores, where the required kinds of commodities 
are offered. The probability to decide for
a certain store as next destination will be the greater, the more
of the required commodities are offered there, the greater 
the {\sl assortment} is, the lower
the {\sl price level} is, and the shorter the {\sl way} to the store is. 
\par
In general, there are several ways to the chosen destination. The probability
to decide for a certain way will decrease with the corresponding distance,
but the readiness for taking detours is growing with the available time.
\par
When arriving at the chosen destination (store), 
the pedestrian will buy a commodity of 
a certain kind. The probability for buying a commodity of a certain kind
during a given time interval
will be increasing with 
the assortment, and with the number
of commodities which are required of this kind. It will be the lower
the higher the price level is. 
These probabilities determine the time 
which is necessary for buying one of the required commodities. 
\par
The purchase
of a required commodity changes the remaining demand and calls for a decision
about the next destination. Since it depends on the distance which destination
is prefered, the same store will usually be chosen again
as long as there is a demand for 
other commodities that are 
offered there (if the prices for
these commodities are not too high, and if the assortment 
with respect to the remaining demand is not too low). 
\par
A pedestrian will leave the shopping area when the demand is satisfied,
i.e. when he or she has bought all of the required commodities.
\par
A detailled model for the {\sl route choice behavior} of pedestrians and its
dependence on their demand has been developed, simulated and 
empirically tested by {\sc Borgers} and {\sc Timmermans} \cite{Borg1,Borg2}.

\subsection{Pedestrian movement}

The motion of pedestrians starts at special {\sl city entry points} like
bus stops, parking lots or metro stations. The choice of a certain entry
points depends on a pedestrian's demand. 
\par
The pedestrian's movement is, then, guided by his or her 
{\sl desired velocity}.
Whereas the {\sl direction} of the desired velocity
is given by the way to the
chosen destination, the {\sl desired speed} of pedestrians is distributed
{\sc Gauss}ian \cite{Rennen,Frauen,Soldaten}. 
The desired speed of pedestrians may
be varying with time. For example, it is increased in the case of 
delays in order to reach a certain destination well-timed.
\par
Since unexpected obstacles and other pedestrians 
have to be avoided, the {\sl actual
velocity} of a pedestrian will normally differ from the desired velocity.
Interactions with other pedestrians are characterized by {\sl avoiding 
maneuvers} and {\sl stopping processes}. They determine the {\sl capacity}
of a pedestrian area. During the interaction free time
pedestrians are accelerating, and trying to approach their desired
velocity again. 
\par
Deviations from the originally chosen way also result from unexpected 
{\sl attractions} like shop windows or entertainment along the 
pedestrian area. Such attractions may lead to {\sl spontaneous stops}
(``impulse stops'').
\par
A detailled model for the movement behavior of pedestrians is given in
\cite{Helbing1}.

\section{Computer simulations}

An ideal method of testing the model described in section \ref{model} is a 
{\sl Monte Carlo simulation} of pedestrian dynamics with a computer.
The results of these simulations can be compared with empirical
data (see \cite{Borg1,Borg2}) 
or with {\sl films} of pedestrian flow. 
\par
Computer simulations can 
be used as a powerful tool for town- and traffic-planning:
They allow to determine an optimal design of pedestrian
areas and an optimal arrangement of store locations, since they take
into account the pedestrian demand, the city {\sl entry points}, 
the {\sl location}
of the stores and the {\sl capacity} of the pedestrian areas.
The capacity depends on the pedestrian density and the pedestrian flows (see
sect. \ref{flow}). It is, therefore, a function of the {\sl size} 
and the {\sl geometry} of a pedestrian area.

\section{Pedestrian groups}

From the behavior of individual pedestrians 
some results concerning
pedestrian groups can be derived. Interesting examples are the
formation of freely-forming groups and the behavior in queues.

\subsection{Formation of freely-forming groups}

Pedestrians who know each other and meet in a pedestrian area
by chance may form a group, and stay
together for a talk. However, a pedestrian will join another pedestrian only,
if the motivation (the attraction) to do so is greater than 
the motivation to get ahead.
The pedestrian will leave the moment at which the 
motivation to join the group becomes less than 
the increasing motivation to get ahead with the
desired velocity (which is growing according to the delay resulting from the
stay). If, right from the beginning, the motivation
to get ahead is greater than the motivation to join a certain person or group
the pedestrian will normally not stop for a talk.
\par
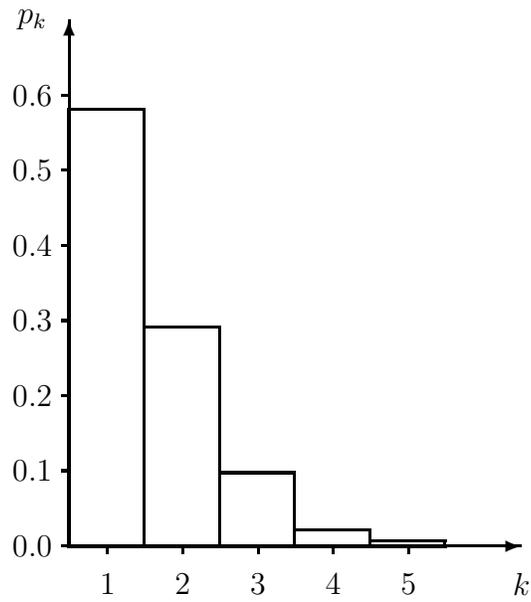
\begin{figure}[htbp]
\unitlength1cm                                                      
\begin{center}
\begin{picture}(6,7)(-0.5,-1)
\thicklines
\put(0,0){\vector(1,0){6}}
\put(-0.01,0){\vector(0,1){7}}
\put(0,0){\framebox(0.97,5.8){}}
\put(1,0){\framebox(0.97,2.9){}}
\put(2,0){\framebox(0.97,0.96){}}                                              
\put(3,0){\framebox(0.97,0.2){}}
\put(4,0){\framebox(0.97,0.05){}}                                  
\put(0.5,-0.5){\makebox(0,0){1}}
\put(1.5,-0.5){\makebox(0,0){2}}                                         
\put(2.5,-0.5){\makebox(0,0){3}}
\put(3.5,-0.5){\makebox(0,0){4}}
\put(4.5,-0.5){\makebox(0,0){5}}
\put(6,-0.5){\makebox(0,0){$k$}}
\put(-0.5,0){\makebox(0,0){0.0}}
\put(-0.5,1){\makebox(0,0){0.1}}
\put(-0.5,2){\makebox(0,0){0.2}}
\put(-0.5,3){\makebox(0,0){0.3}}
\put(-0.5,4){\makebox(0,0){0.4}}
\put(-0.5,5){\makebox(0,0){0.5}}
\put(-0.5,6){\makebox(0,0){0.6}}
\put(-0.5,7){\makebox(0,0){$p_k$}}
\put(0.5,0){\line(0,-1){0.1}}
\put(1.5,0){\line(0,-1){0.1}}
\put(2.5,0){\line(0,-1){0.1}}
\put(3.5,0){\line(0,-1){0.1}}
\put(4.5,0){\line(0,-1){0.1}}
\put(-0.01,0){\line(-1,0){0.1}}
\put(-0.01,1){\line(-1,0){0.1}}
\put(-0.01,2){\line(-1,0){0.1}}
\put(-0.01,3){\line(-1,0){0.1}}
\put(-0.01,4){\line(-1,0){0.1}}
\put(-0.01,5){\line(-1,0){0.1}}                                                 
\put(-0.01,6){\line(-1,0){0.1}}
\end{picture}
\end{center}
\caption[ ]{Group size distribution of freely-forming groups: The 
frequency of groups consisting of $k$ members is given by the truncated
{\sc Poisson} distribution
$p_k = \frac{1}{\mbox{\scriptsize e}^{\lambda} - 1} \frac{\lambda^k}{k!}$
($k=1,2,\dots$) with a situation specific constant $\lambda$.\label{poisson}}
\end{figure}

As a consequence of this joining and leaving behavior, a truncated
{\sc Poisson} distribution results for the group size
(see fig. \ref{poisson}) \cite{Helbing1}.
This has been already
derived and empirically tested by {\sc Coleman} \cite{Col1,Col2}.

\subsection{Behavior in a queue}

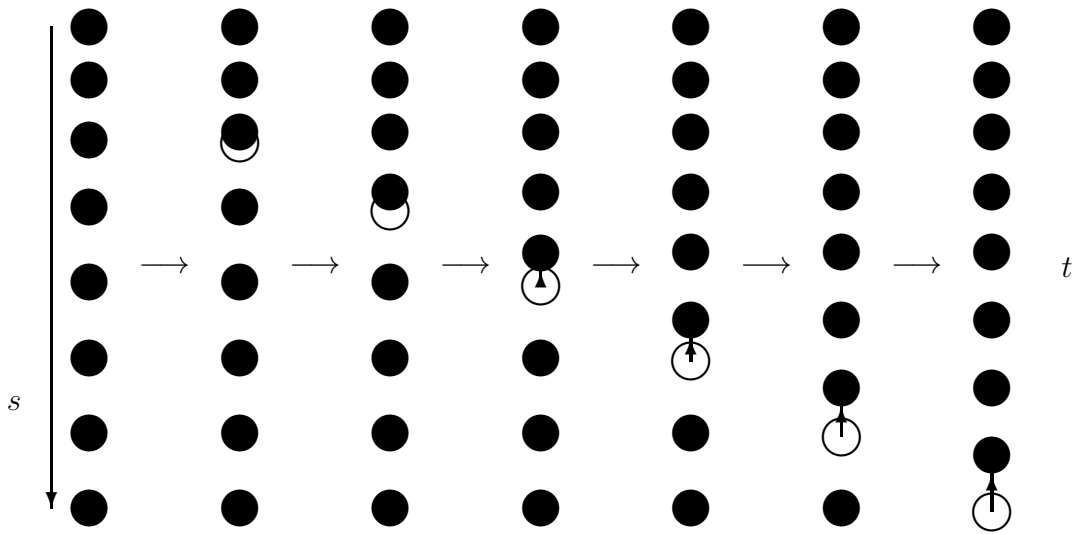
\begin{figure}[htbp]
\unitlength1cm
\thicklines
\begin{center}
\begin{picture}(13.5,7.5)(0,0)
\put(-0.5,7){\vector(0,-1){6.4}}
\put(-1,2){\makebox(0,0){$s$}}
\put(1,3.8){\makebox(0,0){$\longrightarrow$}}
\put(3,3.8){\makebox(0,0){$\longrightarrow$}}
\put(5,3.8){\makebox(0,0){$\longrightarrow$}}
\put(7,3.8){\makebox(0,0){$\longrightarrow$}}
\put(9,3.8){\makebox(0,0){$\longrightarrow$}}
\put(11,3.8){\makebox(0,0){$\longrightarrow$}}
\put(13,3.8){\makebox(0,0){$t$}}
\put(0,7){\circle*{0.5}}
\put(0,6.3){\circle*{0.5}}
\put(0,5.5){\circle*{0.5}}
\put(0,4.6){\circle*{0.5}}
\put(0,3.6){\circle*{0.5}}
\put(0,2.6){\circle*{0.5}}
\put(0,1.6){\circle*{0.5}}
\put(0,0.6){\circle*{0.5}}
\put(2,7){\circle*{0.5}}
\put(2,6.3){\circle*{0.5}}
\put(2,5.45){\circle{0.5}}
\put(2,5.6){\circle*{0.5}}
\put(2,4.6){\circle*{0.5}}
\put(2,3.6){\circle*{0.5}}
\put(2,2.6){\circle*{0.5}}
\put(2,1.6){\circle*{0.5}}
\put(2,0.6){\circle*{0.5}}
\put(4,7){\circle*{0.5}}
\put(4,6.3){\circle*{0.5}}
\put(4,5.6){\circle*{0.5}}
\put(4,4.55){\circle{0.5}}
\put(4,4.8){\circle*{0.5}}
\put(4,3.6){\circle*{0.5}}
\put(4,2.6){\circle*{0.5}}
\put(4,1.6){\circle*{0.5}}
\put(4,0.6){\circle*{0.5}}
\put(6,7){\circle*{0.5}}
\put(6,6.3){\circle*{0.5}}
\put(6,5.6){\circle*{0.5}}
\put(6,4.8){\circle*{0.5}}
\put(6,3.55){\circle{0.5}}
\put(6,4.0){\circle*{0.5}}
\put(6,3.55){\vector(0,1){0.2}}
\put(6,2.6){\circle*{0.5}}
\put(6,1.6){\circle*{0.5}}
\put(6,0.6){\circle*{0.5}}
\put(8,7){\circle*{0.5}}
\put(8,6.3){\circle*{0.5}}
\put(8,5.6){\circle*{0.5}}
\put(8,4.8){\circle*{0.5}}
\put(8,4.0){\circle*{0.5}}
\put(8,2.55){\circle{0.5}}
\put(8,3.1){\circle*{0.5}}
\put(8,2.55){\vector(0,1){0.3}}
\put(8,1.6){\circle*{0.5}}
\put(8,0.6){\circle*{0.5}}
\put(10,7){\circle*{0.5}}
\put(10,6.3){\circle*{0.5}}
\put(10,5.6){\circle*{0.5}}
\put(10,4.8){\circle*{0.5}}
\put(10,4.0){\circle*{0.5}}
\put(10,3.1){\circle*{0.5}}
\put(10,1.55){\circle{0.5}}                        
\put(10,2.2){\circle*{0.5}}
\put(10,1.55){\vector(0,1){0.4}}
\put(10,0.6){\circle*{0.5}}
\put(12,7){\circle*{0.5}}
\put(12,6.3){\circle*{0.5}}
\put(12,5.6){\circle*{0.5}}
\put(12,4.8){\circle*{0.5}}
\put(12,4.0){\circle*{0.5}}
\put(12,3.1){\circle*{0.5}}
\put(12,2.2){\circle*{0.5}}
\put(12,0.55){\circle{0.5}}
\put(12,1.3){\circle*{0.5}}
\put(12,0.55){\vector(0,1){0.5}}
\end{picture}
\end{center}
\caption[]{Behavior in a queue: If an individual moves forward a
little due to the growing pressure of time, the successors
are motivated to do the same. The distance for the
$n$th successor to move forward is usually greater than the one
of the $(n-1)$th successor. The old position of a moving individual
is indicated by an empty circle.\label{queue}}
\end{figure}

If the front of a queue has come to rest, the following phenomenon can
often be observed: After a while, one of the waiting individuals begins to
move forward a little, causing the successors to do the same. This
process propagates in wave-like manner to the end of the queue, and
the distance to move forward increases (see fig. \ref{queue}).
\par

Why do individuals behave in such a paradoxical way?---They do not get
forward any faster but only cause the queue to become more crowded!
The reason is that an individual in a queue keeps a distance, which
corresponds to an equilibrium between the motivation to get ahead 
(the {\sl pressure of time}) and the motivation to keep a certain distance
to the predecessor (the readiness to respect the {\sl territory} 
of the individual in the front).
However, during waiting in the queue, the pressure of time
increases, whereas the territorial effect is time independent. As
a consequence, the individual moves forward a little after a
while.

\section{Pedestrian Crowds}

\subsection{Stochastic formulation}

The mathematical formulation of the model described in section \ref{model} 
leads to a {\sl stochastic equation}, namely the {\sl master equation}.
This equation is extremely complicated: it is impossible to be solved
analytically and hard to be solved 
with a computer. However, from the master equation 
approximate {\sl mean value equations} can be derived, which are
similar to {\sl Boltzmann} equations \cite{Helbing3}. 
These equations can be interpreted
as a {\sl gaskinetic formulation} of pedestrian movement.

\subsection{Gaskinetic formulation}

Before gaskinetic equations have been developed for pedestrian crowds, they
have been already used for the description of {\sl traffic flow} 
\cite{Alberti,Pav,Prig,Pri}.
The gaskinetic formulation of pedestrian behavior
with {\sc Boltzmann}-like equations has
some analogies with the description of ordinary gases, but it takes
into account the effect of pedestrian intentions and interactions.
Some similarities and differences between pedestrian crowds 
and ordinary gases shall be illustrated by two examples:
The behavior an a dance floor on one hand, and the separation of opposite
directions of motion on the other hand.

\subsubsection{Behavior on a dance floor}

\begin{figure}[htbp]
\unitlength1cm
\begin{center}
\begin{picture}(6,6)
\put(0.9,0.8){\circle{0.43}}
\put(0.4,1.1){\circle{0.43}}
\put(1.5,0.5){\circle{0.43}}
\put(0.5,0.4){\circle{0.43}}
\put(1.1,0.1){\circle{0.43}}
\put(1.9,0.1){\circle{0.43}}
\put(2.4,-0.1){\circle{0.43}}
\put(2.8,0.3){\circle{0.43}}
\put(3.3,0){\circle{0.43}}
\put(3.8,-0.2){\circle{0.43}}
\put(4.1,0.3){\circle{0.43}}
\put(4.5,-0.1){\circle{0.43}}
\put(4.8,0.4){\circle{0.43}}
\put(5.4,0.3){\circle{0.43}}
\put(5.1,0.8){\circle{0.43}}
\put(5.5,1.2){\circle{0.43}}
\put(5.8,0.8){\circle{0.43}}
\put(6,1.3){\circle{0.43}}
\put(5.7,1.8){\circle{0.43}}
\put(6,2.3){\circle{0.43}}
\put(6.3,2.8){\circle{0.43}}
\put(5.8,3.2){\circle{0.43}}
\put(6.1,3.6){\circle{0.43}}
\put(5.7,4.1){\circle{0.43}}
\put(6.3,4.1){\circle{0.43}}
\put(6,4.5){\circle{0.43}}
\put(6.1,5){\circle{0.43}}
\put(5.7,5.4){\circle{0.43}}
\put(5.4,4.9){\circle{0.43}}
\put(4.7,5.6){\circle{0.43}}
\put(5.2,5.7){\circle{0.43}}
\put(4.6,6.1){\circle{0.43}}
\put(4.1,6){\circle{0.43}}
\put(3.7,5.7){\circle{0.43}}
\put(3.5,6.3){\circle{0.43}}
\put(3,5.9){\circle{0.43}}
\put(3,6.5){\circle{0.43}}
\put(2.5,6.2){\circle{0.43}}
\put(2.1,5.8){\circle{0.43}}
\put(1.7,6.1){\circle{0.43}}
\put(1.1,5.9){\circle{0.43}}
\put(1.1,5.3){\circle{0.43}}
\put(0.5,5.9){\circle{0.43}}
%
\put(0.6,5.3){\circle{0.43}}
\put(0,5.2){\circle{0.43}}
\put(0.3,4.7){\circle{0.43}}
\put(-0.1,4.3){\circle{0.43}}
\put(0.2,3.9){\circle{0.43}}
\put(0,3.4){\circle{0.43}}
\put(-0.3,3){\circle{0.43}}
\put(0.1,2.5){\circle{0.43}}
\put(-0.3,2.1){\circle{0.43}}
\put(0.3,1.8){\circle{0.43}}
\put(-0.2,1.4){\circle{0.43}}
\put(0,0.8){\circle{0.43}}
%
%
\put(1.5,1.8){\circle*{0.43}}
\put(1.5,1.8){\vector(-1,3){0.2}}
\put(1.5,1.8){\vector(1,-3){0.2}}
\put(2.7,1){\circle*{0.43}}
\put(2.7,1){\vector(1,1){0.43}}
\put(2.7,1){\vector(-1,-1){0.43}}
\put(4,2){\circle*{0.43}}
\put(4,2){\vector(-1,1){0.43}}
\put(4,2){\vector(1,-1){0.43}}
\put(3,3){\circle*{0.43}}
\put(3,3){\vector(2,1){0.6}}
\put(3,3){\vector(-2,-1){0.6}}
\put(2,4.5){\circle*{0.43}}
\put(2,4.5){\vector(-1,2){0.3}}
\put(2,4.5){\vector(1,-2){0.3}}
\put(3.2,4.8){\circle*{0.43}}
\put(3.2,4.8){\vector(1,1){0.43}}
\put(3.2,4.8){\vector(-1,-1){0.43}}
\put(4,4.1){\circle*{0.43}}
\put(4,4.1){\vector(1,0){0.7}}
\put(4,4.1){\vector(-1,0){0.7}}
\put(5.1,3.2){\circle*{0.43}}
\put(5.1,3.2){\vector(0,1){0.7}}
\put(5.1,3.2){\vector(0,-1){0.7}}
\put(1,3.8){\circle*{0.43}}
\put(1,3.8){\vector(1,2){0.3}}
\put(1,3.8){\vector(-1,-2){0.3}}
\end{picture}
\end{center}
\caption[]{Behavior on a dance floor: Dancing individuals 
(filled circles) show a lower density than the individuals
standing around (empty circles), since they intend to move
with a greater velocity variance.\label{dance}}
\end{figure}
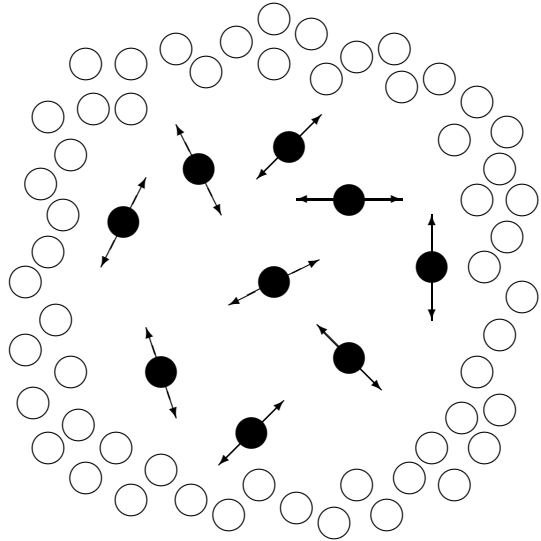

On a dance floor like that of a discotheque, two types of motion can be 
found: One type represents individuals, who want to dance, i.e. intend to move
with a high velocity variance $\theta_d$ (``high temperature''). The second
type represents individuals, who look on the dancers and do not want to
move, i.e. intend to have a low velocity variance $\theta_s$
(``low temperature'').
Dancers and spectators are in equilibrium only, if the mutually exerted
pressure $P= \rho\theta$ of both groups agrees ($P_d = P_s$) \cite{Helbing2}. 
As a consequence, the dancers
are expected to show a lower density $\rho$ than the spectators ($\rho_d <
\rho_s$) (see fig. 
\ref{dance}). This phenomenon can actually be observed.

\subsubsection{Separation of opposite directions of motion}

\begin{figure}[htbp]
\unitlength1cm
\begin{center}
\begin{picture}(10,6)(-0.5,0)
\thicklines
\put(-0.6,-0.2){\line(0,1){6.4}}
\put(3.6,-0.2){\line(0,1){6.4}}
\thinlines
\put(0.5,-0.5){$\Downarrow$}
\put(2.5,6.2){$\Uparrow$}
\put(-1.5,6){(a)}
\put(0,5.6){\circle{0.4}}
\put(0.4,5.2){\circle{0.4}}
\put(-0.3,5.2){\circle{0.4}}
\put(0.3,4.7){\circle{0.4}}
\put(-0.1,4.3){\circle{0.4}}
\put(0.2,3.9){\circle{0.4}}
\put(0.2,3.9){\vector(1,-2){0.25}}
\put(0,3.4){\circle*{0.4}}
\put(0,3.4){\vector(-1,2){0.25}}
\put(-0.3,3){\circle{0.4}}
\put(0.1,2.5){\circle{0.4}}
\put(-0.2,2){\circle{0.4}}
\put(0.3,1.6){\circle{0.4}}
\put(-0.3,1.3){\circle{0.4}}
\put(0,0.8){\circle{0.4}}
\put(-0.3,0.4){\circle{0.4}}
\put(0.4,0.2){\circle{0.4}}
\put(1.1,0.3){\circle{0.4}}
\put(0.8,0.8){\circle{0.4}}
\put(1.1,1.3){\circle*{0.4}}
\put(1.1,1.3){\vector(1,2){0.25}}
\put(0.9,1.8){\circle{0.4}}
\put(0.9,1.8){\vector(-1,-2){0.25}}
\put(0.8,2.4){\circle{0.4}}
\put(1.5,2.8){\circle*{0.4}}
\put(1.5,2.8){\vector(1,2){0.25}}
\put(0.8,3.2){\circle{0.4}}
\put(1,3.7){\circle{0.4}}
\put(0.7,4.1){\circle{0.4}}
\put(1.3,4.2){\circle{0.4}}
\put(1,4.6){\circle{0.4}}
\put(1.2,5){\circle{0.4}}
\put(0.8,5.4){\circle{0.4}}
\put(1.1,5.9){\circle{0.4}}
\put(1.7,5.6){\circle{0.4}}
\put(1.7,5.6){\vector(-1,-2){0.25}}
\put(2.4,5.5){\circle*{0.4}}
\put(1.8,5.1){\circle*{0.4}}
\put(1.8,5.1){\vector(1,2){0.25}}
\put(2.3,4.8){\circle*{0.4}}
\put(1.9,4.3){\circle*{0.4}}
\put(2.2,3.8){\circle*{0.4}}
\put(1.4,3.4){\circle{0.4}}
\put(1.4,3.4){\vector(-1,-2){0.25}}
\put(2.6,2.3){\circle*{0.4}}
\put(2,2.6){\circle*{0.4}}
\put(1.7,2.2){\circle*{0.4}}
\put(2.3,1.7){\circle*{0.4}}
\put(1.9,1.3){\circle*{0.4}}
\put(2.2,0.9){\circle*{0.4}}
\put(1.5,0.8){\circle*{0.4}}
\put(2,0.4){\circle*{0.4}}
\put(1.8,0){\circle*{0.4}}
\put(2.8,0.2){\circle*{0.4}}
\put(3.2,0.6){\circle*{0.4}}
\put(2.7,1.1){\circle{0.4}}
\put(3,1.6){\circle*{0.4}}
\put(3.3,2){\circle*{0.4}}
\put(2.5,2.9){\circle*{0.4}}
\put(3.2,2.7){\circle*{0.4}}
\put(3.1,3.3){\circle*{0.4}}
\put(2.8,3.8){\circle*{0.4}}
\put(3.2,4.3){\circle*{0.4}}
\put(2.7,4.6){\circle*{0.4}}
\put(3,5.3){\circle*{0.4}}
\put(3.3,5.8){\circle*{0.4}}
\put(5,6){(b)}
\thinlines
\put(8,1.8){\circle*{0.54}}
\put(8.1,2.1){\vector(1,2){0.9}}
\dashline{0.2}(7.9,2.1)(7,3.83)
\put(7,3.83){\vector(-1,2){0}}
\put(8,4.2){\circle{0.54}}
\put(7.9,3.9){\vector(-1,-2){0.9}}
\dashline{0.2}(8.1,3.9)(9,2.17)
\put(9,2.17){\vector(1,-2){0}}
\put(6.7,2.2){\makebox(0,0){$p$}}
\put(9.7,2.2){\makebox(0,0){$(1-p)$}}
\put(6.3,3.8){\makebox(0,0){$(1-p)$}}
\put(9.3,3.8){\makebox(0,0){$p$}}
\end{picture}
\end{center}
\caption[]{(a) Opposite directions of motion normally use separate lanes.
Avoiding maneuvers are indicated by arrows.
(b) For pedestrians with an opposite direction of motion it is advantageous,
if both prefer either the right hand side or the left hand side when
trying to pass each other. Otherwise, they would have to stop in order
to avoid a collision. The probability $p$ for choosing the right hand side
is usually greater than the probability $(1-p)$ for choosing the left hand
side.\label{separation}}
\end{figure}
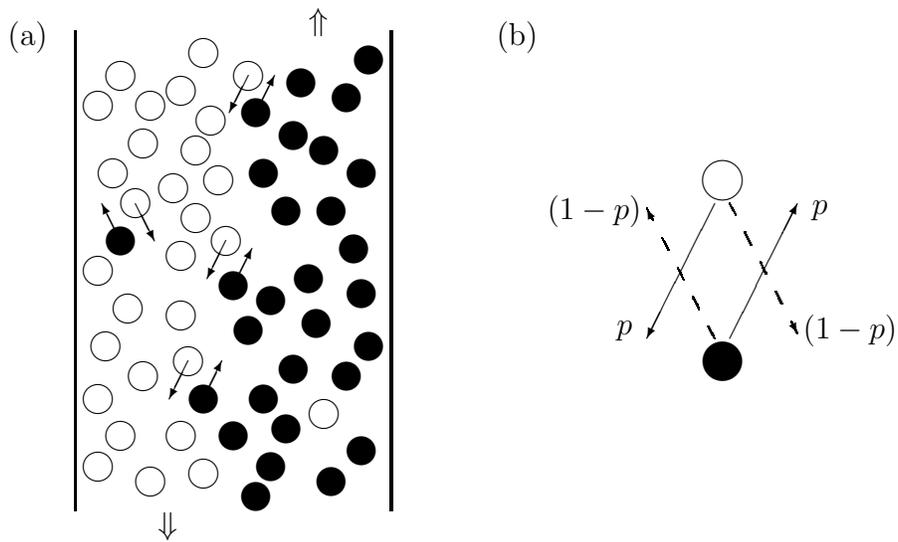

In a street, footway or pedestrian area normally (at least) two opposite
directions of motion (two opposite {\sl flows}) can be found. The interaction
rate (i.e. the rate of avoiding maneuvers and stopping processes) becomes
minimal, if pedestrians with opposite desired directions of motion use
{\sl separate lanes} (see fig. \ref{separation}a). 
Actually, a separation of opposite directions of motion 
arises, if the pedestrian density exceeds a certain value. The width of the
lanes can again be evaluated with the equilibrium condition for
the mutual pressure of both flows \cite{Helbing2}.
\par

A separation of opposite flows and a reduction of stopping processes only
results, if pedestrians with an opposite direction of motion 
both prefer either the
right hand side or the left hand side when trying to pass each other
(see fig. \ref{separation}b). 
In most cases, the right hand side is being prefered,
in spite of the fact that a preference of the left hand side would have the
same effect. This {\sl break of symmetry} can be understood as {\sl
phase transition} \cite{Helbing1,Helbing4}: 
With both sides being originally equivalent,
one side is being prefered by a growing majority, once it has been 
prefered at random.

\subsection{Fluid dynamic equations} \label{flow}

Often, one is only interested in quantities like the density, mean
velocity and velocity variance of pedestrians, but not in the detailled
velocity distribution. In this case, fluid dynamic equations are sufficient.
They can be derived from the gaskinetic equations as mean value equations
for the quantities of interest. This has been explicitly shown in 
\cite{Helbing2}, but investigations on fluid dynamic properties of pedestrian
crowds have been already made by {\sc Henderson} 
\cite{Henderson,Rennen,Frauen,Schiefe}.
The similarity of the motion of pedestrian crowds 
with the motion of ordinary fluids
can be best seen by comparison of quick-motion pictures of pedestrians
with streamlines of fluids. 
Nevertheless, the fluid dynamic equations for pedestrians
contain some additional terms, which take into account the intentions
and interactions of pedestrians.
\par
From the fluiddynamic equations the following conclusions for the
optimization of pedestrian flows can be drawn:
\begin{itemize}
\item Crossings of different directions of motion should be avoided
(if necessary, by bridges, traffic lights or round-about traffic).
\item Opposite directions of motion should use separate lanes.
At a narrow passage, pedestrians should
walk by turns.
\item Great velocity variances should be avoided. This can be done by
walking in formation.
\item If obstacles or narrow passages are unavoidable, they should 
be given an {\sl aerodynamic design}.
\end{itemize}

\section{Conclusions}

Models for pedestrian movement can be developed on several levels.
On the microscopic level one has to describe the individual
behavior, and a stochastic formulation results. However, the microscopic
level of pedestrian movement can be treated easier by Monte Carlo simulations
with computers. 
\par
From the microscopic formulation a gaskinetic formulation
can be derived, which describes the mezoscopic level. On the
macroscopic level, one is confronted with fluiddynamic equations,
which can be derived from the gaskinetic equations. 
\par
All levels
of description take into account pedestrian intentions and interactions.
Models for pedestrian movement can be used as a powerful tool for
town- and traffic-planning.

\end{document}